\documentclass{icrc}

\usepackage{times}
\usepackage{graphicx} 

\begin{document}

\title{Small Scale Anisotropies of UHECRs from Super-Heavy Halo Dark Matter}
\author[1,2]{P. Blasi}
\affil[1]{Osservatorio Astrofisico di Arcetri, Largo E. Fermi 5, 50125
Firenze, Italy}
\author[2]{R. K. Sheth}
\affil[2]{Fermi National Accelerator Laboratory, P.O. Box 500, Batavia, 
IL 60510, USA}

\correspondence{Blasi Pasquale (blasi@arcetri.astro.it)}

\firstpage{1}
\pubyear{2001}


\maketitle

\begin{abstract}
The decay of very heavy metastable relics of the Early Universe can
produce ultra-high energy cosmic rays (UHECRs) in the halo of our own
Galaxy.
In this model, no Greisen-Zatsepin-Kuzmin cutoff is expected
because of the short propagation distances.
We show here that, as a consequence of the hierarchical
build up of the halo, this scenario predicts the existence of
small scale anisotropies in the arrival directions of UHECRs,
in addition to a large scale anisotropy, known from previous 
studies.
We also suggest some other observable consequences of this scenario 
which will be testable with upcoming experiments, as Auger, EUSO and OWL.
\end{abstract}

\section{Introduction}
The production of heavy particles ($m_X\sim 10^{12}-10^{14}$ GeV)
in the early universe has been studied by several authors (see for instance 
\citet{kolb,bere,kuzmin,kt}). If these particles have a sufficiently long
lifetime they can play a role as dark matter candidates and possibly 
generate ultra-high energy cosmic rays, as widely discussed by 
\citet{bere,sarkar,bbv,blasi}. For this to work, these particles must be 
quasistable, which implies lifetimes larger than the present age of the
Universe. Though problematic, this feature can be accomplished,
as discussed by \citet{bere} among others.

This model provides specific predictions for the  spectrum, composition 
and anisotropy of the arrival directions of UHECRs, and these findings
can be compared with present and upcoming data.  
We will concentrate here on the expected small scale anisotropies,
after providing a short summary of the previous studies on the topic.

The decay of the heavy relics is expected to result mainly in the 
generation of hadronic jets initiated by quark and antiquarks. The 
content of the jets is strongly dominated by pions ($95\%$), with 
a small contamination in the form of protons and neutrons (and their
antiparticles). The decay products of the pions are dominated by gamma 
rays (that represent the bulk of ultra-high energy particles 
in this model), neutrinos and electron-positron pairs.

Some features of the hadronization process are sufficiently well 
understood to allow the following predictions:
{\it i)} a flat energy spectrum of UHECRs; {\it ii)} composition dominated 
by gamma rays rather than by protons (see however \citet{sarkar}). 
Moreover, as in all top-down models, 
heavy elements are expected to be completely absent.
Unfortunately present data on the composition is extremely poor and it is
impossible to rule out or confirm the presence of gamma rays in the UHECR
events (however recent evidence against a gamma ray dominated composition
seems to arise from the analysis carried out by \citet{zas}).

\citet{dt} and \citet{bbv} first addressed the issue of the anisotropy, 
recently discussed in detail by \citet{sarkar1}.
The anisotropy results from the asymmetric position of the Earth in 
the Galaxy, so that the flux of UHECRs coming from the direction of 
the galactic center should be appreciably larger than
the flux from the anticenter direction. 
\citet{bm} and \citet{medina} considered this issue
more quantitatively, taking into account the exposure of the present 
experiments.
All authors concur that the present data is consistent with 
the predictions of the relic model for practically all reasonable values  
of the model parameters.

Recently an interesting pattern has arisen from the analysis of the 
events with  energy larger than $4\times10^{19}$ eV: \citet{agasa}
presented a sample with this energy cut, comprising 47 events, whose overall 
distribution in space does not show appreciable deviation from isotropy.  
However, 3 doublets and one triplet could be identified within an 
angular scale of $2.5^o$, comparable with the
angular resolution of the experiment. A complete analysis, 
including the whole set of UHECR events above $4\times 10^{19}$ eV
from the existing experiments was performed by \citet{watson}. 
This extended sample comprises
92 events and shows 12 doublets and two triplets (each triplet is also 
counted as three doublets) within an angle of $3^o$.  
The chance probability of having more than this number of doublets 
was estimated to be $\sim 1.5\%$. Although it is probably too soon to 
rule out the possibility that these multiplets are just a random 
fluctuation, it is instructive to think about the possibility that 
their presence contains some physical information about the sources 
of UHECRs. 
Most of the top-down models for 
UHECRs (e.g. strings, necklaces, vortons, etc.) cannot naturally 
explain the multiplets. 

In the following we will discuss how the multiplets can be interpreted in
the context of the super-heavy dark matter (SHDM) model.

\section{The halo dark matter}

There is mounting evidence that there is plenty of dark matter 
in the universe. The formation of the large scale structures is
relatively well understood in terms of hierarchical clustering,
where larger objects are formed by the continuous merging of 
smaller ones. 

High resolution N-body simulations, e.g., \citet{simul}, suggest that 
the density 
of dark matter particles in galaxy size halos is cuspy in the center, with
the density profile scaling as $\sim r^{-\gamma}$ with $\gamma\sim 1-1.5$ 
on distances $r$ which are much smaller than a core radius, of 
the order of several kpc in size; outside the core, the slope of the 
profile steepens, scaling as $\propto r^{-3}$ at large distances. 

The smooth component of the dark matter can be modeled
in the form suggested by numerical simulations 
(\citet{nfw}):
\begin{equation}
n_H(r)=n^0\frac{(r/r_c)^{-1}}{\left[1+\frac{r}{r_c}\right]^2}
\label{eq:NFW}
\end{equation}
where $r_c$ is the core size and $n^0$ is a normalization parameter. 
These two parameters can be set by requiring that the halo 
contains a given total mass ($M_H$) and that the velocity dispersion 
at some distance from the center is known (in the case of the Galaxy, 
the velocity dispersion is $\sim 200$ km/s in the vicinity of our 
solar system.). 
The parametrization of the dark matter density profile is not 
unique, and there is some discussion in the literature in this respect. 
Alternative fits to the simulated dark matter halos and a discussion of 
whether or not simulated halos appear to be consistent with observations 
are provided by \citet{simul}.

In addition to the smooth dark matter distribution, represented by
eq. (\ref{eq:NFW}), N-body simulations also show that there is a 
clumped component which contains $\sim 10-20\%$ of the total mass. 
The clumps form galactic halos by merging, mainly at about $z\sim 3$
although the process is continuous. Much of the mass initially in a 
small clump which 
falls onto and orbits within the larger halo after that time gets 
tidally stripped from it.  The amount of mass which is lost from 
any given clump increases as the distance of closest approach to 
the galactic halo center decreases; the mass is stripped away, from 
the outside in, as the clump falls towards the center.  
We will call the size of a clump, after its outsides have been stripped 
away, the tidal radius of the clump.  Dynamical friction makes 
the clumps gradually spiral in towards the halo center.

The spatial distribution of the clumps in the halo is not the 
same as that of the dark matter.  We found that a good fit to 
the joint distribution in clump mass and position in the simulations 
of \citet{simul} is
\begin{equation}
n_{cl}(r,m) = n_{cl}^0 \left(\frac{m}{M_H}\right)^{-\alpha} 
\left[1+\left(\frac{r}{r_c^{cl}}\right)^2\right]^{-3/2},
\label{eq:clumps}
\end{equation}
where $n_{cl}^0$ is a normalization constant, $r_c^{cl}$ is the 
core of the clumps distribution, and $\alpha$ describes the relative 
numbers of massive to less massive clumps. The simulations suggest that 
$\alpha\sim 1.9$ (\citet{simul}).  The constraints on the core size are 
weaker---we will study the range where $r_c^{cl}$ is between 3 and 
30 percent of $R_H$. 
In (\citet{simul}), a halo with $M_H\approx 2\times 10^{12}~M_\odot$
contains about $500$ clumps with mass larger than $\sim 10^8~M_\odot$. 
This sets the normalization constant in eq. (\ref{eq:clumps}).

Clumps in the parent NFW halo are truncated at their tidal radii.  
The tidal radius of a clump depends on the clump mass, the density 
profile within the clump, and on how closely to the halo center it 
may have been.  
We assume that clumps of all mass are isothermal spheres (even though 
they are not truly isothermal, \citet{simul} suggest this is reasonably 
accurate):  $\rho_{cl}(r_{cl})\propto 1/r_{cl}^2$, 
where $r_{cl}$ is the radial coordinate measured from the center of 
the clump. The tidal radius of a clump ($R_{cl}$) at a distance $r$ from 
the center of the parent halo is determined by requiring that the 
density in the clump at distance $R_{cl}$ from its center equals the 
local density of the NFW halo at the distance $r$.   
This means that
\begin{equation}
R_{cl} = \left(\frac{m}{4\pi n^0 x_c}\right)^{1/3} x^{1/3} 
\left[ 1+\frac{x}{x_c} \right]^{2/3},
\label{eq:size}
\end{equation}
where $x_c=r_c/R_{H}$, $x=r/R_H$ and $R_H$ is the virial radius of 
the halo, of order $300$ kpc.  The average overdensity within $R_H$ 
is about 200 (\citet{simul}).

\section{Small scale anisotropies in UHECRs}

As shown by \citet{bbv} and \citet{medina}, the total 
(energy integrated) flux of 
UHECRs per unit solid angle from a smooth distribution of dark matter 
particles in the halo is:
\begin{equation}
\frac{d\Phi}{d\Omega} \propto \int_0^{R_{max}} dR\, n_H(r(R)),
\label{eq:flux}
\end{equation}
where $R$ is the distance from the detector, and $r$ is the distance 
from the galactic center (so $R$ and $r$ 
are related by trigonometrical relations accounting for the
off-center position of the Earth in the Galaxy). 
The upper limit, $R_{max}$, depends on the line of sight. 

The existence of a clumped component changes the flux in eq. 
(\ref{eq:flux}) only in that $n_H$ should be replaced with the total 
dark matter density, the sum of the smooth and the clumped components. 

\citet{bereproc} was the first to point out that overdense regions
would generate an excess in the UHECR flux, that would eventually
result in multiplets of events.

To see how important the clumped contribution is, we used two different 
ways of simulating the observed number of events.  
The first approach consisted of calculating the flux
per unit solid angle [eq. (\ref{eq:flux})] along different lines of 
sight directly, taking into account the smooth plus clumped contributions 
to the total density profile [eqs.~\ref{eq:NFW} and~\ref{eq:clumps}]. 
Once a smooth flux map distribution had been obtained, the UHECR events 
were generated from this distribution. 
In the second approach, the events were generated in two steps.  
First, a random subset of the dark matter distribution, which is supposed 
to represent the subset of particles which decayed, was generated.  
The second step was to draw particles from this distribution, and 
then weigh by the probability that the event would actually have been 
detected---so a chosen particle generates an event with probability 
$\propto 1/r^2$, where $r$ is the distance between the particle and 
the detector. 
In both codes, the detector was assumed to be at the position of the
Earth in the Galaxy, and a cut on the directions of arrival was 
introduced to account for the exposure of the AGASA experiment (taken
here as an example).

Fig. 1 shows as an example one of the generated flux maps: 
the map represents
the ratio of the total flux including the contribution from clumps,
to the flux obtained by using a smooth NFW profile.  The various free 
parameters were $r_c=8$ kpc, $r_c^{cl}=10$ kpc, and the mass distribution 
was truncated at a clump mass of $1\%$ of the mass of the NFW halo. 
This sort of plot emphasizes the clump contribution. 

 \begin{figure}[t]
 \vspace*{2.0mm} 
 \includegraphics[width=8.3cm]{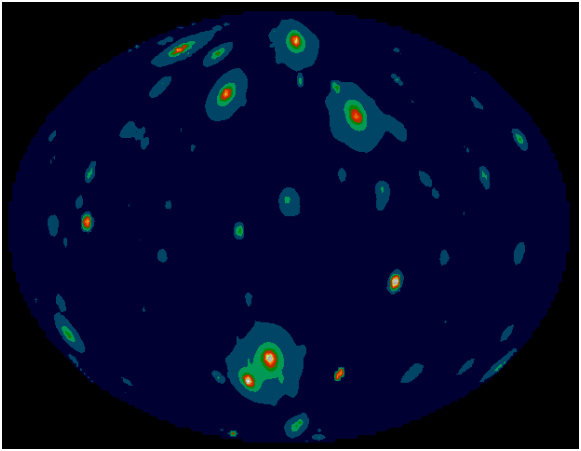} 
 \caption{Small scale anisotropies in the direction of arrival of 
UHECRs from the decay of super-heavy relics.}
 \end{figure}

To calculate the small scale anisotropies, we generated $10^4$ mock 
samples, each of 92 observed events, and counted the number of 
doublets and triplets for angular scales of 3, 4, and 5 degrees. 
Our codes can also be used to check the corresponding numbers for 
the case of isotropic arrival directions (as in \citet{watson}). Two
sets of values of the cores for the NFW and the clumped component
were adopted, one
in which $r_c=8$ kpc and $r_c^{cl}=10$ kpc (case 1) and the other with 
$r_c=r_c^{cl}=20$ kpc (case 2). The observed numbers of doublets within
3, 4, and 5 degrees for an isotropic distribution of arrival directions
are given in \citet{watson} and are 12, 14 and 20 respectively. 
The number of doublets that we obtain in case 1 are 8, 14, and 21
within 3, 4, and 5 degrees respectively. The probability that
the number of doublets equals or exceeds that observed is
$12\%$, $47\%$ and $57\%$ respectively.  This should be compared with 
the $1.5\%$, $13.4\%$ and $15.9\%$ quoted by \citet{watson} for an
isotropic distribution of arrival directions. 

We repeated the same calculation for the case 2. The corresponding 
averages and probabilities of exceeding the observed number of 
doublets within 3, 4 and 5 degree scales are 6.6, 12, and 18, 
and $4.5\%$, $29\%$ and $36\%$ respectively. 

In both cases 1 and 2, the number of doublets on angular scales of 
4 and 5 degrees is consistent with the observed values; presumably 
the discrepancy at 3 degrees is random chance. 

We have also studied the occurence of triplets.  
There is some ambiguity as to how a triplet is best defined; we have 
chosen to define triplets as configurations in which all three pairs 
would have been classified as doublets.  (This means, for example, that 
a co-linear configuration of two doublets is not necessarily a triplet.)   
With this definition, the average number of triplets in case one is 
0.5, 1.5 and 3, with the probability of having more than the 
observed triplets (2, 2, 3 respectively) equal to 4\%, 16\% and 35\%.  
For case 2, the correspondent numbers are 0.4, 1, and 2.5 triplets 
and $2\%$, $8\%$ and $20\%$ for the probabilities to have more triplets
than observed.

What is responsible for the multiplet-events in the SHDM model? 
If we study the case in which all the halo mass is in the smooth 
NFW component, then the number of doublets typically drops by 
one or two.  This suggests that the anisotropy due to our position 
in an NFW halo can result in a number of multiplets of events which 
is considerably larger than if the arrivals were from an isotropic 
background.  The number of multiplets from the clumped component is 
mainly affected by the presence of large nearby clumps, whose number 
depends on the high mass cutoff imposed in the mass function of clumps. 
A maximum mass of $1\%$ of the halo mass implies a total mass in the 
clumps of $\sim 10-15\%$ of $M_H$, 
consistent with the results of the simulations by \citet{simul}. 
Larger cutoffs imply larger mass fractions, 
which are harder to reconcile with the N-body simulations.

The study of the composition, together with an improved 
measure of the spectrum of UHECRs, should nail down the nature of the 
``real'' sources of UHECRs and confirm or rule out the SHDM model.

\section{Conclusions}

In this paper we focused on the small scale anisotropies expected in
the model of super-heavy relics as the sources of UHECRs. In general,
top-down models for the production of particles having energy in excess of
$10^{20}$ eV do not generate small scale anisotropies, with the possible
exception of chance occurrences, that however can be isolated by a 
sufficiently large statistics of events. 
We demonstrated here that the model of super-heavy relics in the halo 
of our galaxy is an exception. This result is due to two main factors: 
1) the inhomogeneous distribution of the smooth component of dark matter
in the halo, and 2) the presence of a clumped component, predicted by 
N-body simulations as a consequence of hierarchical structure formation.

We find that with the current statistics, the number of multiplets 
(doublets and triplets) of events above $4\times 10^{19}$ eV predicted
by the model agrees with the observed one. These multiplets are mainly due 
to the smooth component, and in this sense they are a chance occurrence
determined by the inhomogeneous dark matter distribution. The presence of
the clumps makes the occurrence of multiplets slightly more probable.

With an increased statistics of observed events, a quantity that would better
describe the clustering of events on any scale is the correlation 
function, rather than the number of doublets and triplets, as described 
by \citet{bs00}.
This increased statistics will eventually become available with upcoming
detectors as the Pierre Auger Observatory (\citet{auger}) or future  
experiments as EUSO/Airwatch (http://ifcai.pa.cnr.it/ifcai/euso.html) 
and OWL (http://owl.gsfc.nasa.gov).

The large enhancement in the number of events will also allow for 
a crucial discrimination between the model of super-heavy relics in the
halo and other models, in terms of composition and spectra. In fact the 
super-heavy relic model predicts a gamma ray dominated composition of
the highest energy events and a very flat spectrum, as determined by 
the hadronization process.

\begin{acknowledgements}
This work was partly
supported by the DOE and the NASA grant NAG 5-7092 at Fermilab.
\end{acknowledgements}

\end{document}